\documentclass[epjST]{svjour}

\usepackage{amsmath,amssymb,calc}
\usepackage[bulletsep]{collref}
\usepackage{bbm}
\usepackage[pdftex]{graphicx}
\usepackage{verbatim}
\usepackage{braket}
\usepackage{slashed}
\usepackage{bbold}
\usepackage{xcolor}
\usepackage{mathtools}

\usepackage{hyperref}

\usepackage[final]{feynmf}

\renewcommand{\thesection}{\arabic{section}}

\def\beq{\begin{equation}}
\def\eeq{\end{equation}}

\newcommand{\kcal}{{\mathcal K}}

\newcommand{\beqn}{\begin{eqnarray}}   
\newcommand{\eeqn}{\end{eqnarray}}

\newcommand{\gsim}{\lower.7ex\hbox{$
\;\stackrel{\textstyle>}{\sim}\;$}}
\newcommand{\lsim}{\lower.7ex\hbox{$
\;\stackrel{\textstyle<}{\sim}\;$}}

\numberwithin{equation}{section}

\begin{document}
%

\title{Yang-Mills at Strong vs. Weak Coupling: }
\subtitle{Renormalons,  OPE And All That}
\author{Mikhail Shifman\inst{1}\fnmsep\thanks{\email{shifman@umn.edu}} }

\institute{William I. Fine Theoretical Physics Institute, University of Minnesota, Minneapolis, MN 55455, USA}

%
%
%
%
%
%
%

\abstract{
I discuss various situations in which perturbative expansions are used in Yang-Mills theories with asymptotic freedom
and establish the limits of its applicability.
} 

\maketitle

%

\section{Introduction}

The notion of {\em resurgence} and trans-series associated with it was a breakthrough discovery\,\footnote{For a pedestrian review understandable to physicists (at least, in part) and an exhaustive list of references see \cite{edgar}.} in constructive mathematics in the 1980s mostly associated with the name of Jean Ecalle. 
Resurgence is based on the idea that divergent perturbative series can be made well-defined by invoking additional data on the structure of singularities 
in the coupling constant complex plane plus certain quasiclassical information. 
In this way the standard perturbative series becomes a {\em trans-series},
\begin{eqnarray}
E(g^2)&=&E_{\rm PT, \,\,regularized}(g^2) 
\nonumber\\[3mm]
&+&  \sum_{k=1}^\infty \sum_{l}\sum_{p=0}^\infty 
\underbrace{ \left(\frac{1}{g^{2N+1}}\,\exp\left[-\frac{c} {g^2}\right]\right)^k }_{ \rm k \,\, IA \,\, pairs}\,
{ \left(\log \frac{c}{g^2}\right)^l}\, \underbrace{c_{k, l, p}  g^{2p}}_{\rm regularized\,\,  PT}
\label{trans}
\end{eqnarray}
The role of previously neglected exponential terms of the type $\exp\left(-\tfrac{1}{g^2}\right)$ in the divergent coupling constant expansion  is fully revealed in trans-series
which allow one to achieve any preset accuracy of expansion.

 The method
was scarcely known in modern quantum theory and related areas of physics until 2010s when this construction --  being appropriately adjusted and extended -- was introduced in this context in \cite{au},\,\cite{duun}.
Now it is being further developed by many followers in multiple applications.

The most essential advances have been achieved in such areas as quantum mechanics (QM), partial differential equations, integrable nonlinear systems and orthogonal polynomials. Much more modest is the progress along these lines in asymptotically free quantum field theories, such as Yang-Mills theory, with running constants and complete dynamical restructuring at large distances.

There is a profound difference between the perturbative expansion, say, for the energy eigenvalues of an anharmonic oscillator and
in problems arising in asymptotically free field theories. In the former case a coupling constant is well-defined. It is a small number $g$
which, in principle can be complexified, if necessary, and continued in the complex plane. Perturbative series in $g$
which are usually plagued by factorial divergences in high orders can be made well-defined based on the quasiclassical data obtained in the complex plane.
In this way one arrives at trans-serries, including both regularized perturbation theory and exponential terms $e^{-c/g^2}$ in a systematic manner.

 In Yang-Mills theory there is no dimensionless coupling constant to form a perturbative expansion in the strict sense of this word. If we ignore quarks for the time being then the only parameters of the theory are the dynamical scale $\Lambda$ and the vacuum angle 
$\theta$.\footnote{From experiment we know that $\theta\leq 10^{-10}$ and is irrelevant in phenomenology. However, it plays a role in theoretical studies of Yang-Mills dynamics at strong coupling.} The only expansion parameter appearing in the 't Hooft limit \cite{thlimit} is $1/N$ where $N$ is the number of colors. Quantitative methods for construction of perturbative series in $1/N$  have not yet been developed although a number of qualitative observations exist. 
If such a series could be obtained, say, for the mass of the lightest glueball, we would arrive at
\beq
M_{\rm glueball} = \Lambda\,\sum_{j=0}^\infty \frac{c_j}{N^{k_j}}
\label{1one}
\eeq
where $c_j$ are purely numerical dimensionless coefficients depending only on the quantum numbers of the glueball under consideration. Needless to say, 
the question of convergence in (\ref{1one}) at large $j$ will arise. I think exponential terms of the type $\sim \exp({-cN})$ do appear, but  at the moment it would be prudent to refrain from more detailed speculations.

Passing from Yang-Mills to QCD with massless quarks aggravates the situation. As is well-known, spontaneous breaking of the continuous chiral  symmetry ($\chi$SB)
is not seen in perturbation theory in the gauge coupling, whatever this coupling might mean. Nobody in one's right mind would write that, say, the $\rho$ meson mass can be extracted from a perturbative series even amended by additional quasiclassical information. 
However, there is a range of questions in which QCD perturbation theory is widely used. 

In QCD and similar theories it is quite common to add external sources to use them as tools. For instance, a virtual photon produces a quark-antiquark pair eventually evolving in a cascade of hadrons. If the momentum $Q$ injected in this way can be translated in  Euclidean  we acquire a large external parameter $Q/\Lambda$ and can develop\,\footnote{In the 't Hooft limit $N\to \infty$, in addition to external momenta $Q\gg\Lambda$, other external parameters of a similar nature can enter the game. For instance, if we consider a set of ``radial excitations" of a meson with the given quantum numbers, the number of excitation $k$ at large $k$ can be used as an expansion parameter. }
 a perturbation theory in $\alpha_s (Q)\sim \left(\log Q/\Lambda\right)^{-1}$.\footnote{I use $\alpha_s=\tfrac{g^2}{4\pi} $ with the subscript $s$ in strongly coupled theories while $\alpha$ without the subscript is reserved for weakly coupled theories.}
In a large number of problems (e.g moments of the structure functions in deep inelastic scattering) people learned how to build expansions in $\alpha_s (Q)$ up to three or even four loops. These results can be compared with experiment. With time such  comparisons becomes more and more accurate.

I hasten to add that even in the problems with a large parameter $Q/\Lambda$ the $\alpha_s (Q)$ expansion cannot be made closed -- i.e. cannot be continued to any desirable accuracy, as is the case in QM trans-series. 
The problem lies deeper than just the existence of $\sim k!$ graphs with $k$ loops.  Were this the only obstacle, it could be 
dealt with either through resurgence or by passing to the 't Hooft limit  \cite{thlimit}. In this limit only planar graphs survive, and the number of the planar graphs does not grow factorially \cite{KNN}.\footnote{As I have mentioned above, at large but finite $N$ a factorial divergence will most probably reappear in the $1/N$ series.}

More serious issues exist.
Any physical observable is measured in Minkowski space, not in Euclidean. Conversion from a Euclidean calculation to Minkow\-ski predictions 
 is controllable in a limited sense at best.
Second, the $\alpha_s (Q)$ expansion is intrinsically ill-defined \cite{hooft} because even if 
$\alpha_s (Q)\ll 1$ any  Feynman diagram saturated at $p\sim Q$  (I stress, {\em any}) still contains contributions
from virtual momenta $p\sim \Lambda$. In this domain the coupling $\alpha_s$ is not defined, simply because the Lagrangian formulated in terms of quarks and gluons ceases to exist. 

Therefore,  Feynman graphs are in fact represented by a complicated combination:  the $\alpha_s (Q)$ series,
(i.e. a series of terms of the type $\sum_k \left\{\log \left(Q/\Lambda\right)\right\}^{-k} $ which may also contain loglog's, etc.), exponential terms of the type
 $\sum_k \left(\Lambda/Q\right)^k$, coming from IR contributions,
and conversion factors from Euclidean to Minkowski of the type
$\exp\left(-Q/\Lambda\right)^\gamma$. Both power and exponential terms also are generated by the UV physics, namely tails of the UV renormalons and small-size instantons.  All the expansions above depend on $N$. For conversion factors from Euclidean to Minkowski  this dependence is most dramatic and so are deviations 
from perturbative expansions. I am aware of only one example when this conversion is smooth (see Section \ref{secsmooth}).

Can we amend the $\alpha_s (Q)$ series transforming it in a trans-series, such as in Eq. (\ref{trans}) by invoking quasiclassical information?

The resurgence and trans-series program (such as in quantum mechanics)
can {\em not} be fully successful  in QCD-like theories at strong coupling. Underlying dynamics in confining theories at large distances in no way reduces to 
expansion in $\alpha_s$ even being supplemented by additional quasiclassic analyses. Confinement of the Nambu-Mandelstam-'t Hooft type in the SUSY setting 
was demonstrated  \cite{seiberg1},\cite{seiberg2}  to emerge from the dual Mei{\ss}ner effect -- a very special non-perturbative
feature of the Yang-Mills vacuum -- and so is $\chi$SB. Both are crucial at distances $\gg \Lambda^{-1}$ and leave no trace in perturbation theory.

Summarizing, if one has an external large parameter such as the momentum transfer $Q^2$ or $r(S_1)^{-1}$, (here $r(S_1)$ is the radius of the compactified dimension) one can  classify various expansions:  (i) logarithmic, i.e.  in powers of 
$1/\log(Q^2/\Lambda^2)$ and loglog's/$\log(Q^2/\Lambda^2)$; (ii) in powers of \beq \Lambda^2/Q^2\,\, {\rm or} \,\,\left(\Lambda^2/Q^2 \right)\left(\log(Q^2/\Lambda^2)\right)^\gamma\,;\label{pt}\eeq and, finally, possible exponential effects of the type $\exp\left(-Q/\Lambda\right)^\gamma$. The series in power terms (\ref{pt}) is very likely to be divergent \cite{Shifman1995}.

In the absence of the solution of strong coupling Yang-Mills theories in 4D, the best one can do is the OPE in the form described in detail in the reviews \cite{Shifman_12} (see also references therein) which requires a new delimiting parameter $\mu$,
$$\Lambda\ll\mu\ll Q\,.$$
The latter drops out from all measurable quantities. Introduction of the above parameter  eliminates IR renormalons. The conspiracy between perturbative series and purely gluon operators in OPE  \cite{Parisi1978} remains
valid.

However, if we force the coupling constant to stop running in the infrared (IR) by e.g. Higgsing the gauge theory or formulating it on
$R_3\times S_1$ with $r(S_1)\ll \Lambda^{-1}$ situation changes to mostly quasiclassical and the study of perturbative expansions and isolating non-perturbative terms
may turn out productive and give us certain insights.

Putting Yang-Mills on
$R_3\times S_1$
with $r(S_1)\ll \Lambda^{-1}$   
is one of the most popular techniques exploited today. It was pioneered by \"Unsal et al. \cite{Unsal3}. In this set-up, the coupling constant ceases running at the scale
$ \sim r^{-1}\gg\Lambda$, i.e.
at weak coupling. 
I will return to this set-up later, starting from the most obvious option in which we stay in $R_4$ but Higgs the gauge theory at hand.

\section{Yang-Mills at weak coupling} 

More exactly, I should say ``gauge theories {\em mostly} at weak coupling." Even if all gauge bosons are Higgsed, nontrivial large-distance effects may show up. The most well known example is two-dimensional Higgsed QED in which linear confinement is still present \cite{2DH} (see also \cite{myt}). A less known example is the phenomenon of the baryon number nonconservation at high energies in the Standard Model (see reviews \cite{MattisRub} and references therein, also \cite{myt}, Section 22). At low energies $E/m_{W} \sim 1$ this cross section is of the order of $\exp\left(-2S_{\rm inst}\right)$ where $S_{\rm inst}= \tfrac{8\pi^2}{g^2}$ is the instanton action. As the energy grows the above cross section grows exponentially
until it reaches its maximum $\sim \exp\left(- S_{\rm inst}\right)$ at the energy around the sphaleron mass, 
\beq
M_{\rm sph} = {\rm const} \times {m_W} \frac{2\pi}{\alpha(m_W)}
\label{23}
\eeq
where the numerical constant is of the order of 1.\footnote{This phenomenon is called {\em premature unitarization} \cite{MMS}.} The difference between $\exp\left(-2S_{\rm inst}\right)$ and $\exp\left(-S_{\rm inst}\right)$ is enormous, 90 orders of magnitude or so. 
What is sphaleron? It is a configuration in the space of fields which corresponds to the top of the barrier separating distinct instanton pre-vacua. The latter have the Chern-Simons numbers ${\mathcal K}=0, \pm1, \pm 2,...$ while the sphaleron Chern-Simons numbers are half-integer.
Further details are discussed in Section \ref{higgs}.

The typical number of $W$ bosons and Higgs bosons produced in this process scales as 
$1/\alpha(m_W)$.
The exact maximal value of the {\em  baryon number nonconservation} cross section\,\footnote{Baryon number violation at high temperatures, of the order of $M_{\rm sph}$, can be treated quasiclassically, though \cite{temp}.}
 of the type (for one generation)
\beq
1\,\, {\rm leptons} + 1\,\, {\rm quark} \to 2\,\, {\rm antiquarks} + 1/\alpha_W \,\, {\rm bosons}
\eeq
cannot be obtained in perturbation theory, nor by its quasiclassical extension.  
Even the logarithm of this cross section at $E\sim M_{\rm sph}$  in fact contains an unknown constant of the order of 1 (although less than 2).

However, such examples are rather exotic. Barring such exotic situations, the vast majority of phenomena in Higgsed Yang-Mills or Yang-Mills theories on a small-radius cylinder can be treated semiclassically.

\subsection{Higgsing}
\label{higgs}

An example known to everybody is the SU(2) sector of the Standard Model. If we put the Weinberg angle $\theta_W=0$, all three gauge bosons have the same masses.
I will refer to them as $W$ bosons and assume that $m_W\gg \Lambda$. The maximal value of the running coupling constant is $\alpha_{\max} = \alpha({m_W}) \ll1$. Its running towards the IR is frozen at $m_W$.

Under the circumstances there are no IR renormalons in this theory since $$\left(\beta_1\alpha (Q^2) /4\pi\right) \log (Q^2 /k^2)$$ in Eq. (\ref{eightp}) never approaches unity due to the fact that effectively $k^2\gsim m_W^2$
(see Fig. \ref {bub} and Section \ref{renorm} below). The instanton configuration becomes well defined. Its contribution to low-energy quantities is of the order of 
\beq
\int d^4 x\,  \Lambda^{43/6}_{{\rm SU}(2)}\,v^{-19/6}\,.
\label{34}
\eeq
The overall nonperturbative effect in the  vacuum is an extremely rarified instanton gas. Only if we  approach energies of the order of (\ref{23})
will the strong instanton-antiinstanton interaction manifest itself as was discussed above.
Considering the rarified instanton gas  we could say that it represents  the unit operator or any other purely bosonic operator. The  difference is only in the pre-exponent. The $\mu$ evolution below $m_W$ is frozen and we can descend all the way down to $\mu=m_W$. The situation changes, however, if we add massless (or very light compared to $m_W$) fermions.

The instanton-induced interaction contains all doublets of left-handed SM fermions, say, for the first generation
it has the form
\beq
\left(qqq\right) \left(\ell\right) e^{-S_{\rm inst}} \,.
\label{25}
\eeq
The fermion pre-factor conserves the electric charge and $B-L$. However, obviously it violates the baryon charge conservation by one unit.
The operator appearing in (\ref{25}) is four-fermion and, generally speaking, does not reduce to the unit operator.

\begin{figure}
\centerline{\includegraphics[width=7cm]{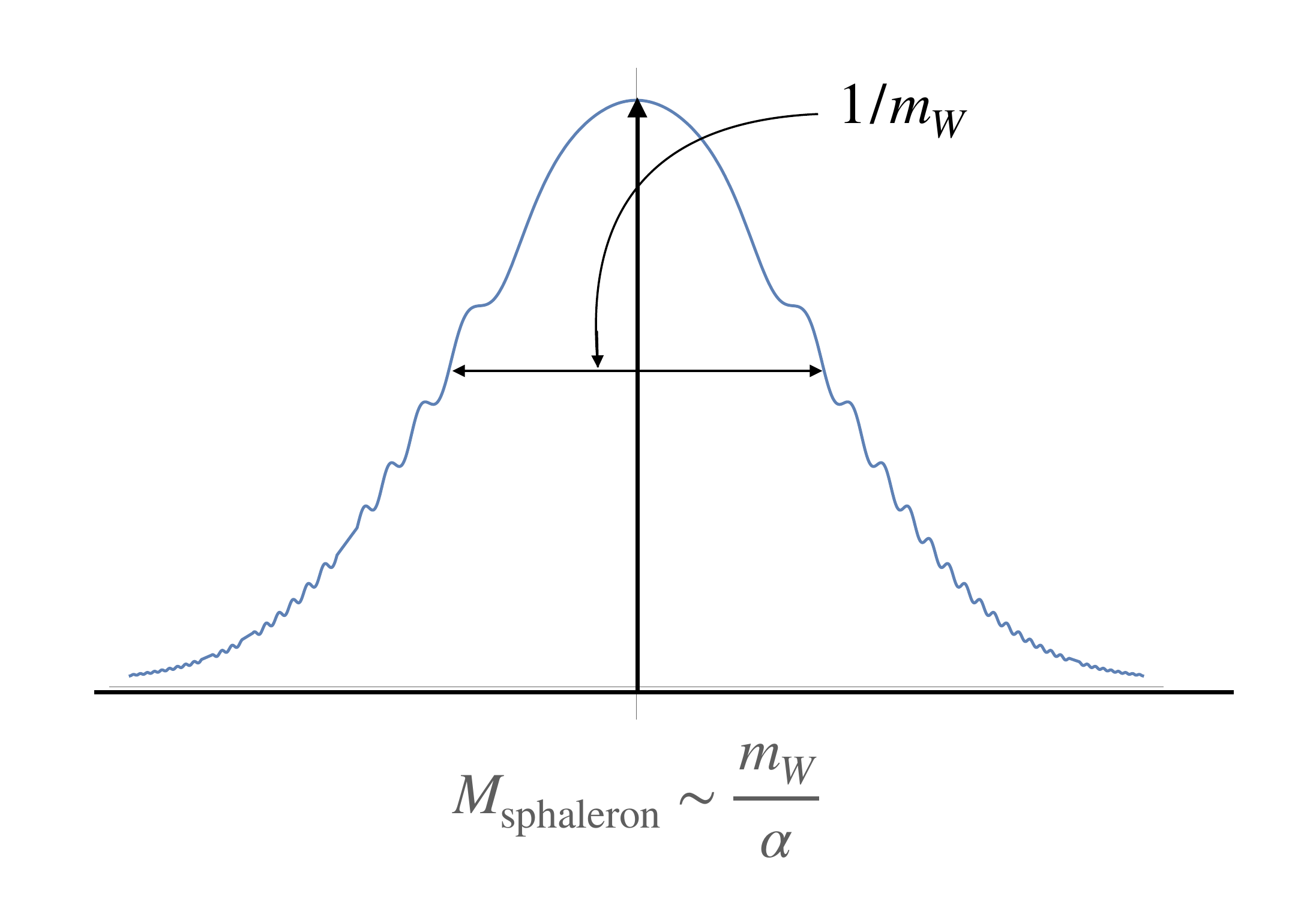}} \caption{The sphaleron is a coherent state of $\sim 1/\alpha$ $W$ bosons. It represents the height of the barrier separating gauge equivalent pre-vacua in Yang-Mills theory.} 
\label{sph}
\end{figure}

 The instanton effects considered above are obtained in Euclidean calculations. Instantons are Euclidean objects. Nevertheless each non-perturbative effect seen in Euclidean must have a clear-cut reflection in Minkowski physics. And, indeed, they do have a reflection
 in the spectrum of the theory. In addition to the $W$-boson triplet and the Higgs scalar this theory has
an unstable soliton called a {\em sphaleron} of a huge mass (\ref{23}) and a typical size $\sim m_W^{-1}$ (see Fig. \ref{sph}). The ratio $M_{\rm sph}/m_W\gg 1$. Such a relation between characteristic parameters is typical for quasiclassical objects. 
The physical meaning of the sphaleron is the height of the top of a barrier in the functional space \cite{klink},\cite{DD} separating pre-vacua with the Chern-Simons charges $\kcal = \kcal^\prime \pm 1$. The sphaleron  is a coherent state of $\sim 1/ \alpha(m_W)$ $W$ bosons. The sphaleron and the associated phenomenon of exponential growth of the $W$-boson production at energies below the sphaleron mass is indirectly related to
the Euclidean instanton (see \cite{myt}, Sections 21.14 and 22,  and references therein).

In the 't Hooft limit, the sphaleron mass (which scales as $N$ if $m_W $ is fixed) becomes infinite and disappears from the spectrum; simultaneously, the instanton contribution vanishes as $e^{-N}$.

Another non-perturbative effect surviving in the Higgsed Yang-Mills  theory is that of the UV renormalon. There is no new physics in it, see Section \ref{six}.
The $g^2$ series coming from the UV renormalon is summable (although not expandable in a regular $g^2$ series).

Long ago Fradkin and Shenker argued \cite{FS} that the following continuity takes place.\footnote{ Fradkin and Shenker argument referred to Yang-Mills on the lattice;
later it was extended.}
{Suppose that, in addition to gauge fields, say in the SU(N) group, a given
non-Abelian theory contains a set of Higgs fields in the fundamental representation, which, by
developing vacuum expectation values (VEVs)  completely Higgs the gauge
group. The theory is at weak coupling provided all gauge bosons are heavy.

Then, upon decreasing all the above VEVs
in proportion to each other from large to small values triggering a strong coupling regime 
we do {\em not}
pass through a Higgs-confinement phase transition. Rather, a
crossover from weak to strong coupling takes place. 

One can argue that this is the case in many different ways. Perhaps
the most straightforward line of reasoning is as follows. Using the
Higgs field in the fundamental representation one can build
gauge-invariant interpolating operators for \textit{all} possible
physical states. They span the space of all possible global
quantum numbers (such as spin and discrete symmetries of the Lagrangian).

 The K\"{a}llen--Lehmann spectral functions
corresponding to these operators, which carry complete information
on the spectrum, depend smoothly on $v$. When the latter parameter
is large the Higgs description is more
convenient; when it is small it is more convenient to think in terms
of bound states. There is no sharp boundary and no phase transition.

This seems quite surprising -- the Higgs regime does not look even remotely similar to that occurring at  strong coupling. And, nevertheless, the passage from weak to strong coupling proceeds without phase transition.\footnote{However, if we add massless quarks the $\chi$SB phase transition can occur.}
} The same phenomenon takes place on a cylinder under certain conditions.

\subsection{Yang-Mills on a cylinder}

As was suggested in \cite{Unsal3}, if one considers Yang-Mills theory on $R_3\times S_1$ rather than on $R_4$ under certain conditions
the theory becomes weakly coupled, with no phase transition on the way from weak to strong coupling. This is called {\em adiabatic continuity}. The most important condition is conservation of the center symmetry which is equivalent to a vanishing Polyakov line. In Yang-Mills theory on $R_3\times S_1$ one can force 
the Polyakov line to vanish either by adding massless adjoint quarks (one is enough) or through a double trace deformation.
As we continuously change $r(S_1)$ from very small (weak coupling) to very large (strong coupling)
 there is no phase transition, in much the same way as in the example of Section \ref{higgs}. The change of the regimes proceeds through a crossover. 
 Qualitative regularities established at weak coupling are also applicable to strong coupling.
 In pure Yang-Mills 
on $R_3\times S_1$  the center symmetry is spontaneously broken at small $r(S_1)$ and therefore the confining  limit   $r(S_1)\to \infty$ with restored center symmetry 
is separated by a phase transition. 

The compactified dimension is assumed to be spatial, $r(S_1)\ll \left(N\Lambda\right)^{-1}$, with judiciously chosen 
boundary conditions on fermion fields (in most cases periodic, sometimes twisted). Thus, the boundary conditions are non-thermal. 
Then, the SU$(N)$ gauge group is Higgsed, SU$(N) \to $U$(1)^{N-1}$, all gauge bosons {\em not} belonging to the Cartan subalgebra acquire masses
$m_W\gg \Lambda$.  
 
 \vspace{2mm}
 
The advantages of this method of the IR regularization are as follows:

(i) One can consider $N$ as a free parameter, in particular, explore the limit $N\to\infty$.
The 't Hooft limit (large-$N$) is much easier to implement on $R_3\times S_1$ than through Higgsing on $R_4$. In this limit  only planar diagrams survive; their number does {\em not} grow factorially 
\footnote{The number $\nu$ of graphs of genus $h$ with $k$ vertices grows at large $k$ as
$\nu_h(k) \sim (k)^{-b_h} \exp( c k ) $. A sphere (relevant to the planar graphs) has genus 0. The dependence on genus resides only in the index in the coefficient $b_h$ in the pre-exponent.}
with $\ell$ where $\ell$ is the number of loops
\cite{KNN}. 

In what follows I will not dwell on this interesting aspect, referring the reader to the original papers.

(ii) The gauge coupling constant is frozen at $g^2(r(S_1)^{-2})$. To guarantee the weak coupling one must choose
$r(S_1) \ll \Lambda^{-1}$ (remember,  $N$ is not a large parameter in my considerations. Otherwise, one should choose $r(S_1) \ll (\Lambda N)^{-1}$). This allows one to carry out a quasiclassical analysis. 
The class of relevant quasiclassical solutions on $R_3\times S_1$ is {\em richer} than on $R_4$. In addition to instanton-monopoles there are bions \cite{Unsal2007} which  carry both topological and magnetic charges.

(iii) The action of these objects is $N$ times smaller than that of the instanton. 
More exactly,
\beq
S_{\rm frac}= \frac k N\,S_{\rm inst}= k\frac{8\pi^2}{g_0^2\,N}\, 
\eeq
where $k=1$ for instanton-monopole, $k=2$ for bions and so on. That's why sometimes they are referred to as fractons. Say, in pure Yang-Mills theory the instanton-monopole contribution to the vacuum energy density
is proportional to
\beq
\Delta{\mathcal E}_{\rm vac} \sim M_0^4 \exp\left[
-\frac{8\pi^2}{g_0^2\,N} \left(1-\tfrac{1}{12} \right)^{-1}
\right]
\eeq
where $-\tfrac{1}{12}$ in the parentheses presents the contribution of non-zero modes. The above contribution does not vanish at large $N$.

(iv) If we add a massless fermion $\psi$, instanton-monopoles produce a bifermion condensate $\bar\psi\psi$. If we add a number of massless fermions $N_f>1$,
the global flavor SU($N_f$) symmetry can be explicitly broken by twisted boundary conditions along the compact dimension. This can extend adiabatic continuity to include Yang-Mills with massless quarks.

\vspace{1mm}

Renormalons are absent provided $r(S_1)$ is sufficiently small. Nonperturbative effects come from fractons, vacuum expectation values of various local operators, UV renormalons, 
and (at finite $N$) from a factorially large number of the multiloop diagrams. The above effects are conceptually similar to those discussed in Section \ref{higgs}.

%
%
%
%

\section{Renormalons}
\label{renorm}

The issue of renormalons was raised by 't Hooft \cite{hooft} (see also\,\footnote{In fact, Lautrup discovered what would now be called UV renormalons appering in QED due to the Landau pole, in his analysis of $g-2$.}
 \cite{Lautrup,Parisi}, detailed reviews can be found in \cite{rev1}, \cite{rev2} and \cite{Shifman_12}). My point of view is that in Yang-Mills theory this problem is misformulated by many. To explain it I will first outline the original idea and then try to present it from the modern standpoint on confining dynamics.
 
 The authors of \cite{hooft,Lautrup} observed that a special class of the so-called bubble diagrams lead to a factorial divergence of the coefficients of $(g^2)^k$ at large $k$. Then the Borel summation of the above formal series leads to ambiguities.
 
\begin{figure}[h]
   \centerline{\includegraphics[width=8cm]{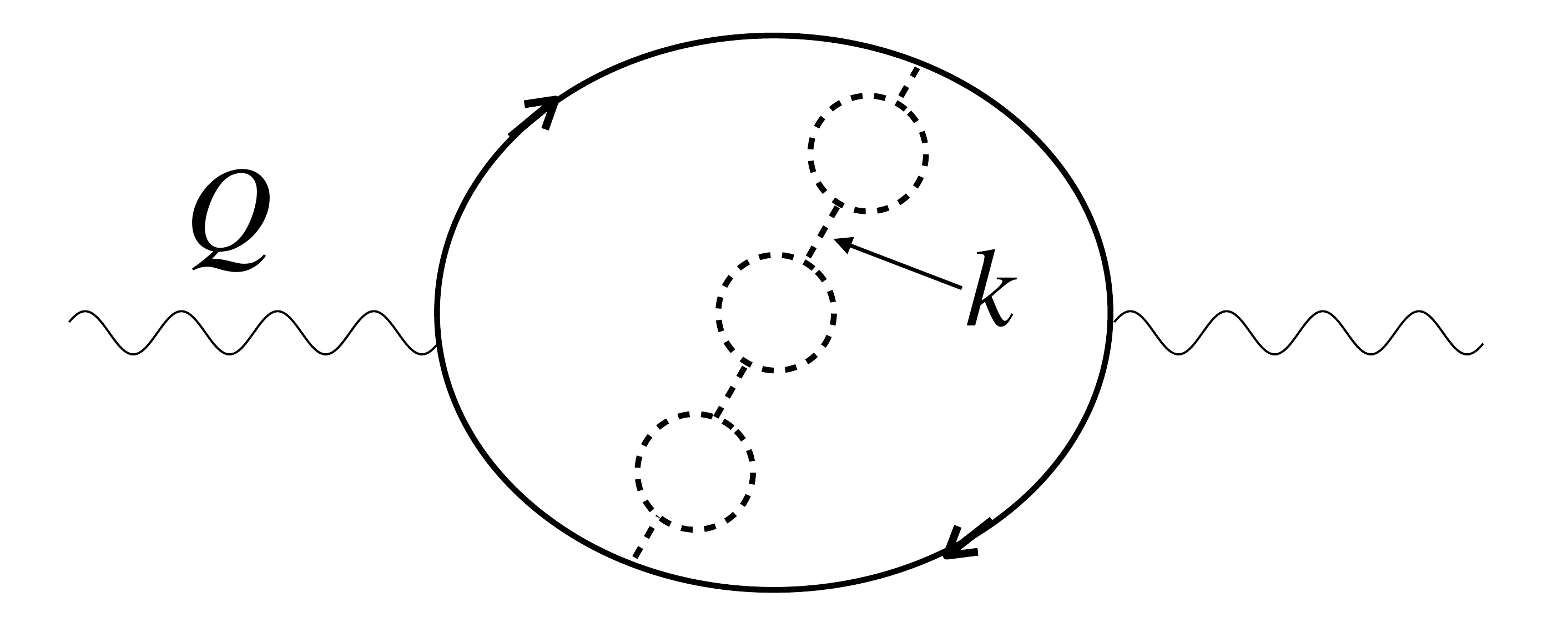}}
\caption{\small The bubble-chain diagrams for the Adler function $D$ representing renormalons. 
Wavy lines stand for an external "photon". Solid lines denote quark propagators, while dashed lines are for gluons.
The quark bubbles are also to be added to the gluon bubbles.
\label{bub}}
\end{figure}
 After one integrates over the loop momentum of the ``large"
fermion loop and the angles of the gluon momentum $k$ in Fig. \ref{bub} one arrives at
\beq
D\propto \int\frac{dk^2}{k^2}\,F(k^2)\, \alpha_s(k^2)
\label{54}
\eeq 
where the function $F(k^2)$ calculated in \cite{neubert} has the limits
\beq
F(k^2) \to \left\{ 
\begin{array}{l}
{{\rm c}_{1}} \left( k^4/Q^4\right),\qquad k^2\ll Q^2\,,\\[2mm]
{\rm c}_2\,  Q^2 \left(k^2\right)^{-1} \log  k^2 , \qquad k^2\gg Q^2\,,
\end{array}
\right.
\label{55}
\eeq

\vspace{2mm}
\noindent
and to the leading order
\beq 
\alpha (k^2) = \alpha (Q^2)
\left[ 1-\frac{\beta_1\alpha (Q^2)}{4\pi} \,
\log (Q^2/k^2)
\right]^{-1} \,.
\label{eightp}
\eeq
Here $\beta_1$ is the first coefficient of the $\beta$ function, $\beta_1= \frac{11}{3}N -\frac 2 3 N_f$.
The upper line in (\ref{55}) is relevant to the IR renormalon, while the lower line to the UV renormalon which will be briefly discussed later.

The confining regime at large distances in Yang-Mills theory at strong coupling implies that Eq. (\ref{eightp}) below is totally inappropriate at
$k^2\lsim \Lambda^2$ since $\alpha (k^2) $ is not defined.

The 't Hooft renormalon analysis is a {\em formal} procedure completely neglecting the above circumstance. Indeed, people formally expand Eq. (\ref{eightp})
 in $\alpha (Q^2)$ and then integrate over $k^2$ from {\em zero} (I emphasize, \underline{zero}) to $Q^2$ ,
   \beq
  D (Q^2)\propto \frac{1}{Q^4} \, \sum_{n=0}^\infty \left(\frac{\beta_1\alpha}{4\pi} 
  \right)^n  \int_0^{Q^2}\,dk^2 \,k^2 \left(\ln \frac{Q^2}{k^2}
  \right)^n\,,\qquad
  \alpha \equiv \alpha(Q^2),
  \label{ninea}
  \eeq
which can be identically rewritten as
\beq
 D (Q^2)\propto  \frac{1}{2}\,\sum_{n=0}^\infty  \left(\frac{\beta_0\alpha}{8\pi} 
  \right)^n  \int_0^\infty\,dy \,
    y^n\, e^{-y}\,,\qquad
 y = 2 \ln \frac{Q^2}{k^2}\,.
  \label{ninepppp}
  \eeq
  Note that $k^2\sim 0$ corresponds to $y \sim\infty$. Formally, the integral in (\ref{ninepppp}) is $n!$
  which makes the sum in (\ref{ninea}) divergent. It is quite obvious, however, that the integral in (\ref{ninea}) cannot run from $k^2=0$ just because Eq. (\ref{eightp}) is defined only provided 
  \beq
  \frac{\beta_1\alpha (Q^2)}{4\pi} \,
\log (Q^2/k^2) < 1
  \eeq
which in turn requires  
\beq
k^2 > \mu^2\,,\qquad \mu^2 = c\Lambda^2\,,\quad c\gg 1\,.
\eeq
  The IR cut off of this type is implemented in {\em bona fide} QCD OPE, see e.g. the review papers \cite{Shifman_12}. Then the factorial growth of the coefficients
  is cut off at a critical value
  \beq
n_* = 2\ln\frac{Q^2}{\mu^2}\,,
\label{r2}
\eeq
and the series in  $\alpha (Q^2)$ in the OPE coefficients is well-defined. (See Fig. 9 in the second paper in Ref. \cite{Shifman_12}.) Alternatively, one can say that the integral in (\ref{54}) is well defined
provided we set the lower limit of integration at $\mu^2\gg \Lambda^2$. Moreover, if we use the exact expression for the function $F(k^2)$ 
we can integrate over $k^2 \in [\mu^2, \infty ]$, the integral will be convergent and well defined.
In addition, it will include the {\bf UV} renormalon in its entirety. However, if we try to expand it in the series in the coupling constant two problems arise:
(i) it is not clear what is exactly the coupling constant we expand in; and (ii) the series will not be convergent, with factorially growing coefficients due to UV renormalons, see below.

In weakly coupled Yang-Mills theories an auxiliary IR cut off $\mu^2$ is not needed. A natural cut off is provided  by either $v^2$ or $\left[ r(S_1)\right]^{-2}$. 
  
  \subsection{A few words about UV renormalons}
  \label{six}
  
In the previous section I emphasized that the IR renormalon is a formal issue provided the OPE is  properly built, including the delimitating parameter $\mu^2\gg\Lambda^2$. At weak coupling no IR renormalons exist to begin with.

UV renormalons do exist at weak coupling since they emerge due to the fact of the slow approach of the gauge coupling $\alpha (k^2)$ to zero as $k^2\to \infty$.
In asymptotically free theories we do~not expect any unknown phenomena in the extreme UV. The coupling constant (\ref{eightp}) is well defined at $k^2\to \infty$.

Usually (see e.g. \cite{rev2}) when discussing UV renormalons people start from (\ref{54}) and the second line in (\ref{55}) and arrive at
  \beq
D (Q^2)\propto {Q^2} \, \sum_{n=0}^\infty \left(\frac{\beta_1\alpha}{4\pi} 
  \right)^n  \int_{Q^2}^\infty\,dk^2\, \frac{1}{k^4}\,(-1)^n \left(\ln \frac{k^2}{Q^2}
  \right)^{n+1}\,.
  \label{612}
  \eeq
The integral in (\ref{612}) is indeed factorial, but sign alternating. It is Borel-summable. However, the whole procedure is totally unreasonable. It is like going from A to B for no reason and then returning back to A. Indeed, instead of expansion (\ref{612}) we should have stared directly from 
  \beq
D (Q^2)\propto {Q^2} \,   \int_{Q^2}^\infty\,dk^2\, \frac{1}{k^4}\,\alpha(k^2) \left(\log k^2 \right)
  \label{613}
  \eeq
where
\beq 
\alpha (k^2) = \alpha (Q^2)
\left[ 1+\frac{\beta_1\alpha (Q^2)}{4\pi} \,
\log (k^2/Q^2)
\right]^{-1} \,.
\label{eightp}
\eeq
The integrand in (\ref{613}) is nonsingular, the integral is convergent but not expandable in the series in the gauge coupling.

\section{Yang-Mills in Banks-Zaks limit at large \boldmath{$N$}}
\label{banksz}

It is instructive to discuss the Banks-Zaks limit \cite{BZ} at large $N$.\footnote{This section resulted from a discussion with M. \"Unsal.} 

At large $N$ only the planar graphs survive and their number does {\em not } grow factorially in high orders \cite{KNN}.  Moreover, instantons
disappear in the 't~Hooft limit. There are no renormalons either.

\begin{figure}
\centerline{\includegraphics{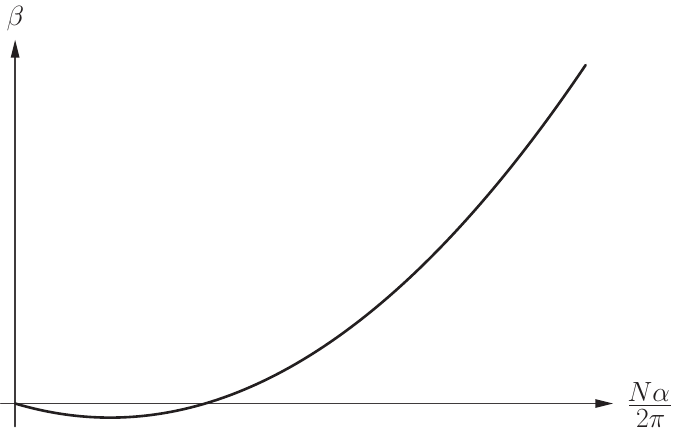}} \caption{The
$\beta$ function at $N_f$ slightly less than $\frac{11}{2} N$. The
horizontal axis presents $a\equiv N\alpha/2\pi$. The zero of the beta
function is at $\frac{8}{75}\nu/N\ll 1$.} \label{ba}
\end{figure}

Indeed, as was noted by Banks and Zaks, at $N_f\lsim \frac {11}{2} N$ we are close to the right edge of the 
conformal window, see Fig. \ref{ba}. Let us have a closer look at the beta function $\beta (a)$,
\beqn
\beta (a) &=& -\beta_1 a^2 -\beta_2 a^3 +..., \qquad a \equiv \frac{N\alpha}{2\pi}\,,
\nonumber\\[3mm]
\beta_{1}&=&\tfrac{11}{3}\left( 1-\tfrac{2}{11}\tfrac{N_{f}}{N}\right),\quad \beta_{2}=\frac{17}{3}-\frac{N_{f}}{6N}\left(13-\tfrac{3}{N^2}\right).
\label{420}
\eeqn
If we choose 
\beq
N_{f}=\tfrac{11}{2}N-\nu,\qquad
0<\nu\ll\tfrac{11}{2}N
\eeq
then the first coefficient, $\beta_1$, is anomalously small,
\begin{equation}
\beta_{1}=\frac 23\,  \frac{\nu}{N}\,,
\eeq
while the second coefficient is of a normal order of magnitude,
\beq
\beta_{2}\approx -\frac{25}{4}
\eeq
and is {\em negative}. This results in an infrared fixed point at a small value of $a$,
\beq
a=a_*= \frac{8}{75}\,\frac{\nu}{N}\,.
\label{424}
\eeq
Since $a_*$ is small, both its value  and the very existence
are reliable. Therefore, if in the UV  we start 
from $a_0 < a_*$ then in the IR the maximal value of the coupling constant is $a_*$ and we flow to the conformal theory with small anomalous dimensions.
The corresponding RG flow is depicted in Fig. \ref{banz}.
\begin{figure}
\centerline{\includegraphics[width=7cm]{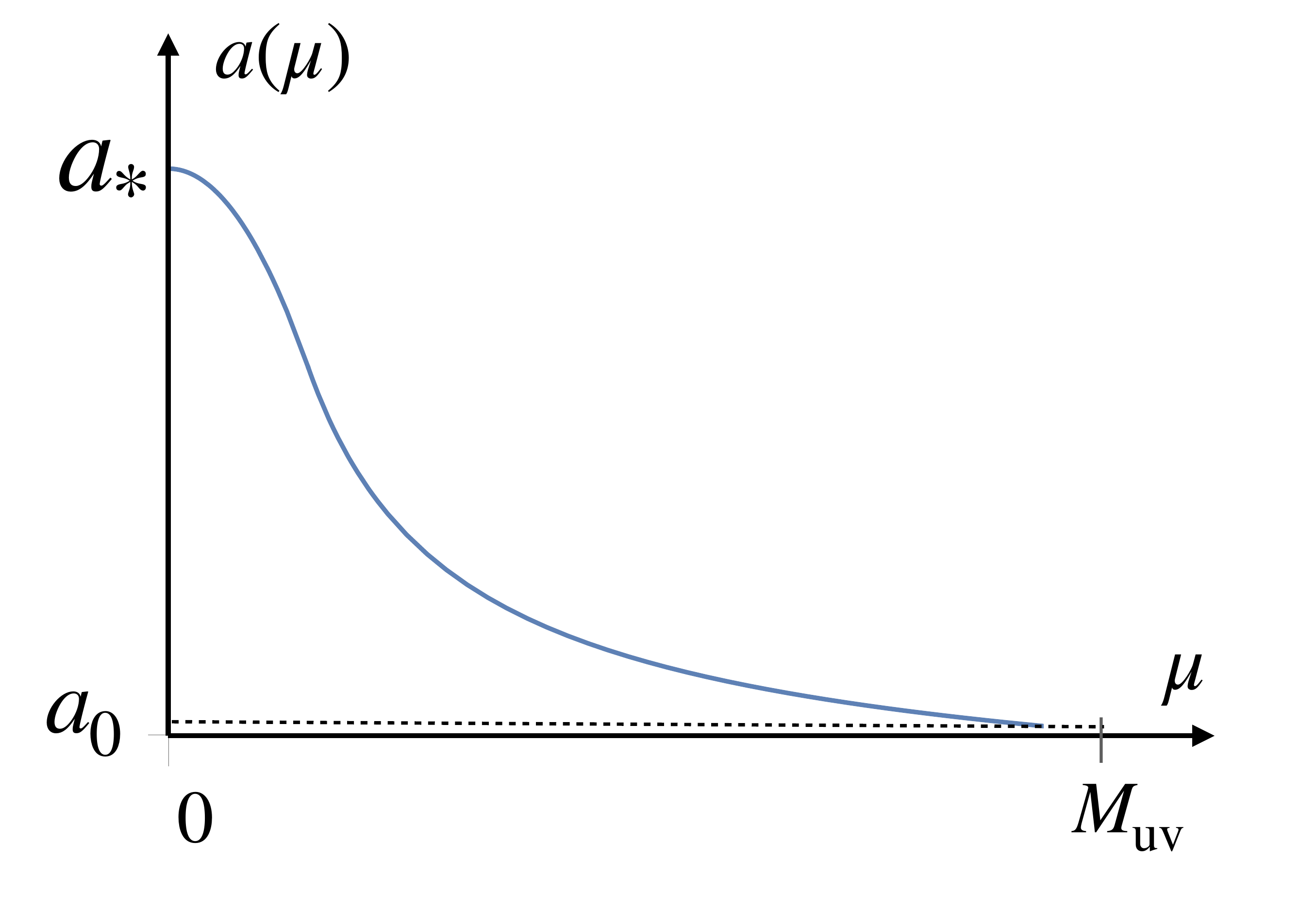}} \caption{The
evolution of $a(\mu)$ from the UV to the IR fixed point at $\mu\to 0$.
The maximal value of $a$ is $a_*\ll1$. } 
\label{banz}
\end{figure}

While the approach to the UV fixed point is logarithmic and is determined by the dynamical scale $\Lambda$, the approach to the IR conformal point is power-like,
\beq
a(\mu ) \to a_* - \left(\frac{\mu}{\tilde\Lambda}\right)^\gamma\,, \qquad \mu\to 0\,.
\label{26}
\eeq
The dimensional parameter $\tilde\Lambda$ is determined also by the value of $a_0=a(M_{\rm uv})$. Typically, $\tilde\Lambda\gg\Lambda$.
The power $\gamma$ in Eq. (\ref{26}) is positive, its numerical value depends on parameters. In the regime discussed above the approach to $a_*$ is from below, see
Fig. \ref{ba}.

The fact that $a(\mu) <\alpha_*\ll 1$ explains the absence of renormalons. Seemingly, there is no source for the factorial divergence
of the perturbation theory coefficients at large order. Nevertheless two dimensional parameters appear in this theory through dimensional transmutation.

If we started from the initial condition $a_0=a(M_{\rm uv})>a_*$ the corresponding regime would dramatically change,  from conformal to Landau, $\alpha (\mu )$ would 
approach $\alpha_*$ from above in the IR, asymptotic freedom would be lost and we  acquire
the Landau growth in the UV.

A remark is in order regarding possible compactification from $R_4$ to $R_3\times S_1$
with $r(S_1)\ll \Lambda^{-1}$. This does not do any good in the discussion of the Banks-Zaks limit. Indeed, the RG flow is interrupted and frozen  at the scale $R^{-1}$ and
we never approach the conformal limit unless we tend $R\to \infty$. In this aspect one can compare the situation with that in fully Higgsed Yang-Mills theory.

Another question to ask is about the emergence of $\tilde \Lambda$ as we approach the conformal regime. Perturbation theory seems well-defined in IR and yet
a dimensional parameter shows up. 
In the conformal regime with anomalous dimensions it should appear in the coefficients in the laws of power fall-off of the correlation functions. 
A related question is what happens to OPE in this regime.  In strongly coupled Yang-Mills it is defined through separation of short and large distances \cite{Shifman_12} which is not applicable near the conformal point. On the other hand the OPE is well formulated in the conformal regime {\em per se}, and even simpler than at strong coupling. 

Needless to say, UV renormalons are present in the theory at hand. They generate factorials in the perturbation series, which is Borel summable, i.e. well defined. 
The remnant of the  Borel summation can be represented by the dynamical scale $\Lambda$.

\section{Smooth passage from Euclidean to 
Minkowski in the conformal window}
\label{secsmooth}

In the Banks-Zaks regime with $a_*\ll 1$ (see (\ref{420}), (\ref{424})) the theory is conformal, with small anomalous dimensions. There is no confinement and no mass gap.
Under these circumstances the quark-gluon description of the theory should be applicable not only in the Euclidean  space but also in Minkowski, with a smooth transition between them. This should be true for any $N$. There are no new thresholds opening as we move in energy 
in ``experiment." 

This issue was discussed in detail in Ref. \cite{adlers} in the framework of ${\mathcal N}=1$ supersymmeric QCD (SQCD). An exact formula was 
derived for the photon-induced Adler $D$ function,
\beq D (Q^2)= \frac{3}{2}  N \sum_f
q_f^2\left[  1 -   \gamma \left(\alpha_s (Q^2) \right)\right]\,,
\label{m5} 
\eeq 
where $f$ is the flavor index, and $q_f$ is the
corresponding electric charge (in the units of $e$). Equation
(\ref{m5}) assumes that all matter fields are in the fundamental
representation of SU$(N)$, although their electric charges can be
different. Above,   $\gamma \left(\alpha_s (Q^2) \right)$ is the anomalous dimension of the matter superfield. It is known only in the form of expansion in 
$\alpha_s (Q^2)$. 

In this section we will study some consequences ensuing \cite{adlers} from the
exact formula (\ref{m5})  in the so-called conformal window (more exactly, close to the right edge of the window,
where the theory is weaky coupled).
 \beq 
 \frac 32 N < N_f
< 3 N\,. 
\eeq

\begin{figure}
\centerline{\includegraphics[width=10cm]{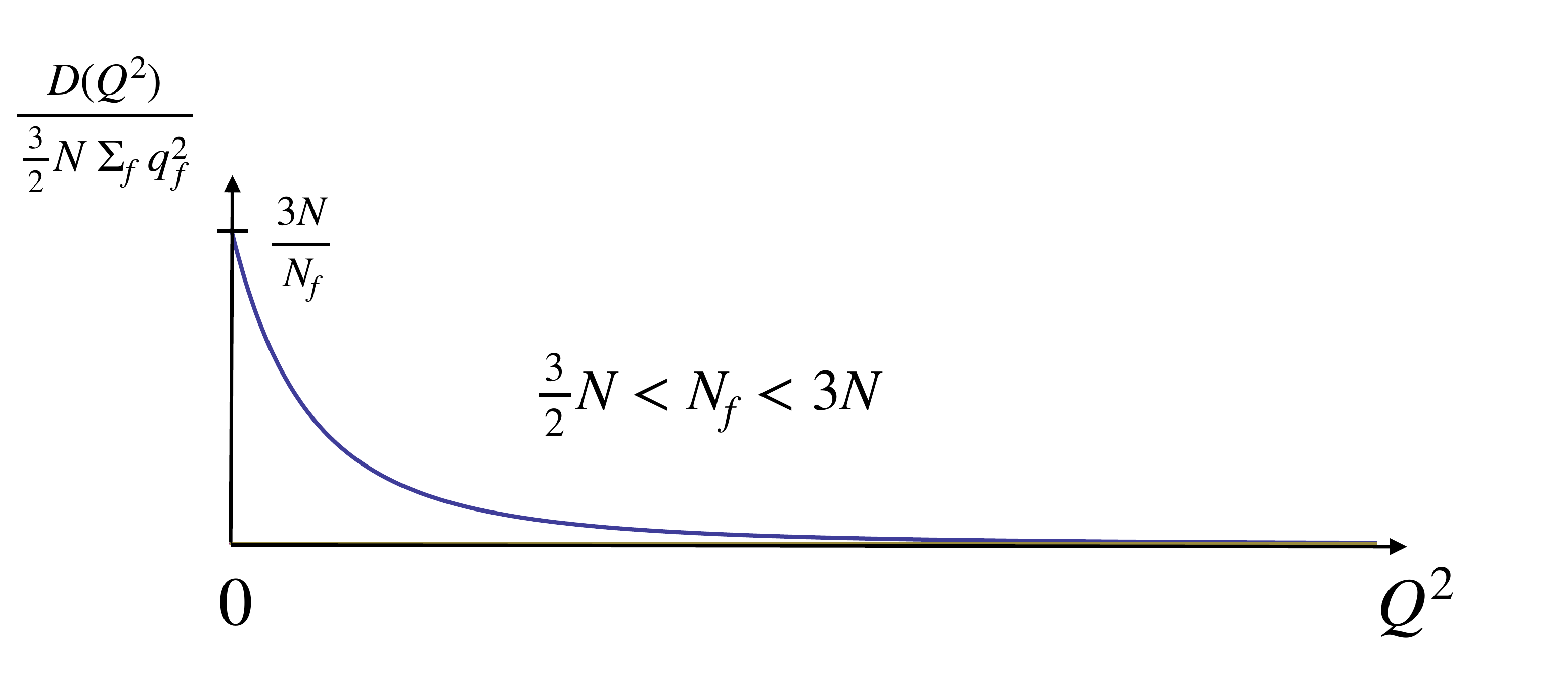}} \caption{\small $D(Q^2)\,\left( \frac 32 N \sum_f
q_f^2\right)^{-1}$ versus $Q^2$ in the conformal window.  The number of co\-lors is $N$  while the number of flavors is $N_f$. The
horizontal axis corresponds to $D(Q^2)\,\left( \frac 32 N \sum_f
q_f^2\right)^{-1} =1$.} 
\label{adle}
\end{figure}

%
Inside this window SQCD flows to the conformal points: $\gamma=0$
in the ultraviolet (asymptotic freedom) and to 
\beq 
\gamma_* =
-\frac{3N-N_f}{N_f}
 \label{82} 
\eeq 
in the infrared. Given $N$ and $N_f$ the maximal value of $\gamma$ is $\gamma_*$. 
The
equation (\ref{m5}) then implies that the Adler function \beq
D(Q^2) \longrightarrow \frac 32 N \sum_f q_f^2    \times
 \left\{
\begin{array}{c}
1\,,\qquad Q^2\to\infty\,,\\[3mm]
\frac{3N}{N_f}\,,\qquad Q^2\to 0\,.
\end{array}
\right.
\label{83}
\eeq
The $Q^2$ evolution of the Adler function in the conformal window is sketched in Fig \ref{adle}.

Of course, we do not know the analytic value of $\gamma \left(\alpha_s (Q^2) \right)$ except in two limiting points 
(\ref{83}). However, we can expect (at least in the 't Hooft limit of  large $N$) that the only source of non-perturbative corrections is the UV renormalon
which smoothly continues from Euclidean to Minkowski so that the predicted curve for 
\begin{equation}
R(s) = \frac{\sigma (e^+e^-\to \mbox{ (s)quarks, gluons(inos)} \to  \mbox{hadrons})}{\sigma
(e^+e^-\to \mu^+\mu^-)}
\end{equation}
obtained from the imaginary part of $D$ is smooth.

\section{Exactly solvable models}
\label{esm}

In special cases of exactly solvable asymptotically free models, the analysis of OPE can be further extended.\footnote{This section is based on discussions with 
Daniel Schubring and Chao-Hsiang Sheu. Detailed results based on these discussions will be published in a separate paper \cite{sss}.}
 Among this class are two-dimensional sigma models, for instance, the O($N$) and CP$(N-1)$ models in the limit of large $N$. Typically we can solve them in the leading and next-to leading order in $1/N$. With the {\em exact} solutions in hand we can first check the $\mu$ independence of OPE \cite{novik} and after this is done answer the question what happens in the limit $\mu\to 0$.  

 The question was raised long ago \cite{david} and recently discussed more than once by \"Unsal et al. in various models. Here we will briefly review  the $O(N)$ 
 model  in which we will discuss the limit of $\mu\to 0$. The most challenging case of the {\em supersymmetric} $O(N)$ sigma model is presented in \cite{sss}.
 
Since one knows the exact answer for all correlation functions, one can {\em define} the ``coupling constant" for perturbation theory,
\beq
g(p^2) \stackrel{\rm def}{=} \frac{4\pi}{N\log\tfrac{p^2}{\Lambda^2}} \,,\qquad \Lambda^2\equiv m^2 = M_0^2\exp\left( -\frac{4\pi}{Ng_0}\right).
\label{526}
\eeq 
At $p^2\gg\Lambda^2$ the definition in (\ref{526}) coincides with the standard perturbative one. 

Dependence on $p/m$ in the exact solution appears in a two-fold way. The exact formula 
contains logarithms of the type $\log p^2/m^2$ and powers of $m^2/p^2$. Of course, in mathematical sense $m^2/p^2$ is just the exponent of $\log p^2/m^2$.
However, it is instructive to keep a double expansion
\beq
\langle n^a(-p)\, n^a(p)\rangle =\sum_{j,\ell} C_{j\ell} \left(\frac{1}{\log p^2/m^2}\right)^j \left(\frac{m^2}{p^2}\right)^{\tfrac{d_\ell}{2}}
\label{211}
\eeq
where the first factor represents coefficients while the second matrix element in the limit $\mu\to 0$.

Now we can make use of the knowledge of the exact Green's functions
in this model \cite{novik},\cite{davis},\cite{sss}. For instance, for the two-point function of the
$n$ fields we have
\beqn
&& \langle n^a(-p), \, n^a(p) \rangle = \frac{N}{p^2+m^2}  +\frac{8m^2\pi}{(p^2+m^2)^2}\int\frac{d^2 k}{4\pi^2}\frac{1}{k^2+4m^2}
\nonumber\\[2mm]
&-&\frac{2}{p^2+m^2}\int\frac{d^2 k}{4\pi^2}\frac{1}{\left(k^2+4m^2\right) J(k^2) \left[\left(k+p\right)^2+m^2\right]}
\nonumber\\[2mm]
&=& \frac{1}{p^2+m^2}\Bigg[N+\frac{2m^2}{p^2+m^2}\log\frac{M_0^2+4m^2}{4m^2}-\left(\tfrac 12 -\gamma+\log \frac{Ng(p)}{4\pi} \right)\nonumber\\
&-&2 \sum_{j=1}^\infty \,(2j)!\,\zeta(2j+1) \left(\frac{Ng(p)}{4\pi}\right)^{2j+1} \Bigg]\nonumber\\
\label{527}
\eeqn
where 
\beq
J(k^2) = \frac{1}{2\pi\sqrt{k^2(k^2+4m^2)}}\log{\left(\frac{\sqrt{k^2+4m^2}+\sqrt{k^2}}{\sqrt{k^2+4m^2}-\sqrt{k^2}}\right)}
\eeq
and after integration in the second line we performed a (formal) expansion in $1/\log\tfrac{p^2}{\Lambda^2}$, see Eq. (\ref{526}). Moreover, $\zeta (s)$ in (\ref{527})
is the $\zeta$ Riemann function and $\gamma$ is the Euler's constant. The expansion over $1/\log\tfrac{p^2}{\Lambda^2}$ can be viewed
as an exact contribution to the unit operator. 

Now, the result (\ref{527}) can be also expanded in $m^2/p^2$. The latter expansion can be viewed as coming from higher dimension operators. Let us limit ourselves by
the identity and dimension-2 operators in OPE. The latter is $\left(\partial_\alpha n^a\right)^2$. In the leading order in  $1/N$
\beq
\langle\left(\partial_\alpha n^a\right) \left(\partial_\alpha n^a\right)\rangle  = -\frac{m^2}{g} +O(1/N)\,.
\eeq
We will consider the subleading in $1/N$ order in this matrix element  below. The OPE takes the form
\beq
\langle n^a(-p), \, n^a(p) \rangle = \frac{N}{p^2}\,  C_{\textbf I} \,{\textbf I} + C_ {(\partial  n)^2}\,\langle\left(\partial_\alpha n^a\right)^2 \rangle
\eeq
where
\beqn
&& 
C_{\textbf I} (p^2) =1 - \frac{1}{N}\left(... +2 \sum_{j=1}^\infty \,(2j)!\,\zeta(2j+1) \left(\frac{Ng(p)}{4\pi}\right)^{2j+1} \right)
\\[4mm]
&& C_ {(\partial n)^2}= \frac{1}{p^4} + \mathcal{O}(1/N)\,.
\eeqn

\vspace{2mm}

\noindent
Both $C_{\textbf I} $ and $\langle(\partial n)^2\rangle$ have formal (purely imaginary) ambiguities associated with the Borel summation of the series at the order $O(1/N)$.
The coefficient $C_{\textbf I} $ formally has an ambiguity 
\beq
\delta C_{\textbf I}  = \mp i\pi \frac{m^{2}}{p^2} \cdot \frac{1}{N}\,.
\eeq
The matrix element of the operator $ \left(\partial_\alpha n^a\right) \left(\partial_\alpha n^a\right)$ also has a purely imaginary ambiguity,
\beq
\langle\left(\partial_\alpha n^a\right)^2 \rangle =
\pm i\pi m^{2}
\eeq
If we assemble them together we observe perfect cancellation (to the given order in  $1/N$). The emergence of purely imaginary ambiguities is the price we have to pay
for our unwillingness to explicitly  introduce the sliding (delimitating) parameter $\mu^2\gg\Lambda^2$, where $\Lambda^2$ is defined in (\ref{526}). Were $\mu^2$ introduced we would have no ambiguities, but will have to trace cancellation of $\mu$ in OPE. Both cancellations are manifestation of the conspiracy inherent to OPE.

I started Section \ref{esm} with the statement:
``With the {\em exact} solutions in hand we can first check the $\mu$ independence of OPE \cite{novik} and after this is done answer the question what happens in the limit $\mu\to 0$." Unfortunately, today we still cannot answer the last part of this  question in QCD and similar theories which operate in the strong coupling regime and are not exactly solvable.. 

\section{Conclusions}

Various sources of nonperturbative effects in Yang-Mills theories at weak coupling are revisited. Although perturbative expansions are widely used 
in asymptotically free Yang-Mills theories it should be realized that their accuracy is limited even at weak coupling.

\section*{Acknowledgments}
I am grateful to Daniel Schubring, Chao-Hsiang Sheu, and M. \"Unsal for very useful discussions.

This work is supported in part by DOE grant de-sc0011842.

\end{document}